\documentclass[11pt]{amsart}
\usepackage{hyperref}
\def \dd{{\rm d}}

\begin{document}
\title{Electrostatics and confinement in Einstein's unified field
theory}
\author{Salvatore Antoci}%
\address{Dipartimento di Fisica ``A. Volta'' and IPCF of CNR, Pavia, Italy}%
\email{Antoci@fisicavolta.unipv.it}%
\author{Dierck-Ekkehard  Liebscher}%
\address{Astrophysikalisches Institut Potsdam, Potsdam, Germany}%
\email{deliebscher@aip.de}%
\author{Luigi Mihich}%
\address{Dipartimento di Fisica ``A. Volta'', Pavia, Italy}%
\email{Mihich@fisicavolta.unipv.it}%
\maketitle
 Einstein's unified field theory was devised already
in 1925 for unifying gravitation and electromagnetism through the
use of a nonsymmetric fundamental tensor $g_{ik}$ and of a
nonsymmetric affine connection $\Gamma^i_{kl}$; since 1945 it was
intensely pursued by both Einstein and Schr\"odinger, who assumed
it to be a complete field theory, not allowing for
phenomenological source terms. However in 1955 a disappointed
Erwin Schr\"odinger wrote: ``It is a disconcerting situation that
ten years endeavour of competent theorists has not yielded even a
plausible glimpse of Coulomb's law." It occurred to us, through
the study of exact solutions found in the meantime, that
physically relevant results instead appear as soon as sources are
allowed for, like it happens with the sources of gravity in the
general relativity of 1915. We consider here the Hermitian version
of the theory. Let
\begin{equation}\label{1}
g_{ik}=g_{(ik)}+g_{[ik]}, \ \
\Gamma^i_{kl}=\Gamma^i_{(kl)}+\Gamma^i_{[kl]},
\end{equation}
be both Hermitian, i.e. endowed with real symmetric and purely
imaginary antisymmetric parts. We pose
\begin{equation}\label{2}
g^{il}g_{kl}=\delta^i_k, \ \ \mathbf{ g}^{ik}=\sqrt{-g}g^{ik},
\end{equation}
where $g\equiv\det{(g_{ik})}$ is real. Then the original field
equations without sources read:
\begin{eqnarray}\label{3}
g_{ik,l}-g_{nk}\Gamma^n_{il}-g_{in}\Gamma^n_{lk}=0,\\\label{4}
\mathbf{ g}^{[is]}_{~~,s}=0,\\\label{5}
R_{(ik)}(\Gamma)=0,\\\label{6} R_{[[ik],l]}(\Gamma)=0,
\end{eqnarray}
where $R_{ik}(\Gamma)$ is the usual Ricci tensor, Hermitian thanks
to (\ref{3}) and (\ref{4}).

 A class of solutions to these equations was
found\cite{Antoci1987}, that depend intrinsically on three
coordinates. These solutions show that there is merit in appending
sources at the right-hand sides of all the field equations, while
preserving the Hermitian symmetry through the symmetrised Ricci
tensor $\bar{R}_{ik}(\Gamma)$ of
Borchsenius\cite{Borchsenius1978}, that is equal to
${R}_{ik}(\Gamma)$ wherever the sources vanish. Works by
Lichnerowicz on the Cauchy problem\cite{Lichnerowicz1954} and by
H\'ely\cite{Hely1954} show that the Bianchi identities look
physically meaningful\cite{Antoci1991} when the metric is chosen
to be a symmetric tensor $s^{ik}$ such that
\begin{equation}\label{7}
\sqrt{-s}s^{ik}\equiv\sqrt{-g}g^{(ik)},
\end{equation}
where $s^{il}s_{kl}=\delta^i_k$, and $s\equiv\det{(s_{ik})}$, and
sources are appended at the right-hand sides of
(\ref{3})-(\ref{6}), in the form of a symmetric energy tensor
$T_{ik}$, and of two conserved currents
$4\pi\mathbf{j}^i=\mathbf{g}^{[is]}_{~~,s}$ and $8\pi
K_{ikl}=\bar{R}_{[[ik],l]}$. Then the Bianchi identities read:
\begin{equation}\label{8}
{\mathbf{T}}^{ls}_{;s}=\frac{1}{2}s^{lk}
\left({\mathbf{j}}^i\bar{R}_{[ki]}
+K_{iks}\mathbf{g}^{[si]}\right).
\end{equation}
The fundamental tensor for the general electrostatic
solution\cite{ALM2005} reads
\begin{equation}\label{9}
 g_{ik}=\left(\begin{array}{rrrr}
 -1 &  0 &  0 & a \\
  0 & -1 &  0 & b \\
  0 &  0 & -1 & c \\
 -a & -b & -c & ~d
\end{array}\right), \ {\text{where:}}
\end{equation}
\begin{eqnarray}\nonumber
d=1+a^2+b^2+c^2, \\\label{10}
a=i\chi_{,x}, b=i\chi_{,y}, c=i\chi_{,z}, \\\nonumber
\chi_{,xx}+\chi_{,yy}+\chi_{,zz}=0.
\end{eqnarray}
The imaginary part $g_{[ik]}$ of the fundamental tensor
looks like the general electrostatic
solution of Maxwell's theory, because the ``potential'' $\chi$
must obey Laplace's equation, and
$$\mathbf{g}^{[is]}_{~~,s}=0\ ,\ \ g_{[[ik],l]}=0,$$
happen to be satisfied. The squared interval reads:
\begin{equation}\label{11}
\dd s^2=s_{ik}\dd x^{i}\dd x^{k} =-\sqrt{d}\left(\dd x^2 +\dd
y^2+\dd z^2-\dd t^2\right) -\frac{1}{\sqrt{d}}(\dd\chi)^2.
\end{equation}
If only one point charge $h$ is present in the ``Bild\-raum'' $x$,
$y$, $z$, $t$, say, at the origin of the coordinates, the
``potential'' $\chi$ is
\begin{equation}\label{12}
\chi=-\frac{h}{(x^2+y^2+z^2)^{1/2}}.
\end{equation}
For this occurrence to happen, a net charge must appear at the
right-hand side of the equation $\mathbf{ g}^{[is]}_{~~,s}=0$.
 The surface $\dd \chi=0$, $d=0$ is the inner border of the
manifold. In the metric sense, this border is a point with a
spherically symmetric neighbourhood.

 A solution for $n$ point charges at equilibrium can be
 built\cite{ALM2005}
by considering, in the ``Bild\-raum'', $n$ closed surfaces
possessing net charges, whose charge distribution reproduces the
one occurring on $n$ conductors at rest, due to their mutual
induction. By changing the shapes and the positions of the
surfaces in order to obtain that both $\dd \chi=0$ and $d=0$ on
each of them, one gets, in the metric sense, $n$ point charges
whose infinitesimal neighbourhood is spherically symmetric. When
this occurs, even approximately, the position of the charges
mimics the equilibrium positions prescribed by Coulomb's
law\cite{ALM2005}.

Axially symmetric solutions allowing for $n$ charged poles
at the right-hand side of equation $R_{[[ik],l]}=0$ are easily
found\cite{ALM2006}. In cylindrical coordinates $x^1=r$, $x^2=z$,
$x^3=\varphi$, $x^4=t$ one writes the fundamental tensor:
\begin{equation}\label{13}
g_{ik}=\left(\begin{array}{rrrr}
  -1 & 0 & \delta & 0 \\
  0 & -1 & \varepsilon & 0 \\
  -\delta & -\varepsilon & \zeta & \tau \\
  0 & 0 & -\tau & ~~1
\end{array}\right),
\end{equation}
\begin{eqnarray}\nonumber
\zeta=-r^2+\delta^2+\varepsilon^2-\tau^2, \\\label{14}
\delta=ir^2\psi_{,r},
\varepsilon=ir^2\psi_{,z}, \tau=-ir^2\psi_{,t}, \\\nonumber
\psi_{,rr}+\frac{\psi_{,r}}{r}+\psi_{,zz}-\psi_{,tt}=0.
\end{eqnarray}
The square of its line element reads:
\begin{equation}\label{15}
\dd s^2=s_{ik}\dd x^{i}\dd x^{k} =\frac{\sqrt{-\zeta}}{r}
\left(-\dd r^2-\dd z^2-r^2\dd {\varphi}^2+\dd
t^2\right)+\frac{r^3}{\sqrt{-\zeta}}(\dd\psi)^2.
\end{equation}
A static solution with $n$ aligned poles built with $K_{ikl}$ is found by
requiring:
\begin{equation}\label{16}
\psi=-\sum_{q=1}^n K_q\ln\frac{p_q+z-z_q}{r}, \ \
p_q=[r^2+(z-z_q)^2]^{1/2};
\end{equation}
$K_q$ and $z_q$ are constants.
 Let us consider the particular case when $n=3$, and $K_1=K_3=K, \
K_2=-K$, with $z_1<z_2<z_3$.\par The charges are always pointlike
in the metric sense; moreover, with the choice shown above, the
metric happens to be spherically symmetric severally in the
infinitesimal neighbourhood of each of the charges\cite{ALM2006}.

  If chosen in this way, the three ``magnetic''
charges are always in equilibrium, like it would happen if they
would interact mutually with forces independent of distance.
  The same conclusion was
already drawn by Treder in 1957 from approximate
calculations\cite{Treder1957}, while looking for electromagnetism
in the theory. In 1980 Treder reinterpreted\cite{Treder1980} his result as
accounting for the confinement of quarks: in the Hermitian theory
two ``magnetic'' poles with unlike signs are confined entities,
because they are permanently bound by central forces of constant
strength.

\end{document}